# Nonlinear Electrodynamics and Optics of Graphene

**S. A. Mikhailov and N. A. Savostianova**

University of Augsburg, Institute of Physics, Universitätsstr. 1, 86159 Augsburg, Germany
E-mail: sergey.mikhailov@physik.uni-augsburg.de

**Summary:** Graphene is a two-dimensional material with strongly nonlinear electrodynamics and optical properties. We present some of our recent theoretical results on the quantum and non-perturbative quasi-classical theories of nonlinear effects in graphene, influence of substrates on graphene nonlinearities, plasma oscillations in graphene in the nonlinear regime and other effects.

## 1. Introduction

Graphene is a two-dimensional (2D) crystal consisting of a single atomic layer of carbon atoms [1]. Charge carriers in graphene have a linear energy dispersion

$$E_{\pm}(\mathbf{k}) = \pm \hbar v_F |\mathbf{k}|, \quad (1)$$

Fig. 1, and behave as massless Dirac quasi-particles; here $v_F \approx 10^8$ cm/s is the material parameter (Fermi velocity). It was theoretically predicted [2] that the linear dispersion (1) should lead to a strongly nonlinear electrodynamic response of this material. This can be seen indeed from simple physical considerations. In contrast to massive particles, the velocity $\mathbf{v} = \partial E_{\pm}(\mathbf{k})/\partial \mathbf{p} = v_F \mathbf{k}/|\mathbf{k}|$ of electrons with the spectrum (1) is not proportional to the momentum, therefore if an electric field $E_0 \cos(\omega t)$ acts on it, its momentum $\mathbf{p} = \hbar \mathbf{k}$ oscillates as $\sin(\omega t)$ but the velocity $\mathbf{v} \sim \mathrm{sgn}[\sin(\omega t)]$ contains higher frequency harmonics. All other nonlinear effects can evidently also be seen in graphene due to the same reasons.

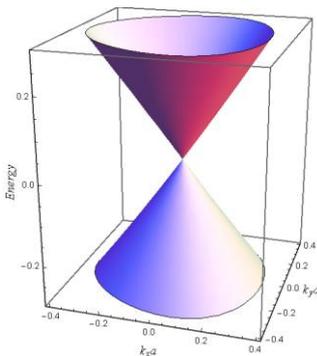

**Fig. 1.** The band structure of graphene electrons near the Dirac points at the corners of the hexagonal Brillouin zone.

This prediction has been confirmed in numerous experiments performed in the last years, in which harmonics generation, four-wave mixing, Kerr effect and other nonlinear phenomena have been observed in graphene at microwave through optical frequencies. In this paper we report about some of our recent results on the theory of nonlinear graphene electrodynamics and optics.

## 2. Theory

### 2.1. Quantum Theory of the Third-Order Effects in Graphene

The simple semiclassical picture of the nonlinear graphene response outlined above is valid at low (microwave, terahertz) frequencies, $\hbar\omega < 2E_F$, when the interband transition from the lower (valence) to upper (conduction) band are not essential. A quantum theory valid at all frequencies was developed in [3]. The third-order conductivity tensor $\sigma^{(3)}_{\alpha\beta\gamma\delta}(\omega_1+\omega_2+\omega_3;\omega_1,\omega_2,\omega_3)$ was analytically calculated within the perturbative approach; here the first argument $\omega_1+\omega_2+\omega_3$ is the output signal frequency which equals the sum of three different input frequencies $\omega_1$, $\omega_2$, and $\omega_3$. Using the function $\sigma^{(3)}_{\alpha\beta\gamma\delta}(\omega_1+\omega_2+\omega_3;\omega_1,\omega_2,\omega_3)$ one can analyze a large number of different nonlinear effects.

*2.1.1. Harmonics generation*

Figure 2 shows the function $\sigma^{(3)}_{xxxx}(3\omega;\omega,\omega,\omega)$ which determines the third harmonic generation from a single graphene layer. Three resonances which are clearly seen at T=0 K and correspond to the one-, two-, and three-photon absorption at the absorption edge ($\hbar\omega=2E_F$, $E_F$ and $2E_F/3$) can be tuned by the gate voltage which controls the Fermi energy and the electron (hole) density in the sample. At room and higher temperatures these resonances get broader (here T is the electron temperature which can essentially exceed the lattice temperature in the nonlinear optics experiments).

*2.1.2. Kerr effect and saturable absorption*

Figure 3 shows the real and imaginary parts of $\sigma^{(3)}_{xxxx}(\omega;\omega,\omega,-\omega)$ at room and a higher temperature. The real part of $\sigma^{(3)}_{xxxx}(\omega;\omega,\omega,-\omega)$ determines the nonlinear absorption in graphene. It is negative at all frequencies, which corresponds to a reduction of the linear absorption in strong electric field, and has a broad resonant feature (at 300 K) around 0.4 eV ($\hbar\omega=2E_F$). The imaginary part of $\sigma_{xxxx}^{(3)}(\omega;\omega,\omega,-\omega)$ is positive at infrared and optical frequencies, has a broad resonance at $\hbar\omega=2E_F$ and changes its sign at low frequencies. The positive sign of Im $\sigma_{xxxx}^{(3)}(\omega;\omega,\omega,-\omega)$ implies an effective negative refractive index of graphene corresponding to the self-defocusing Kerr nonlinearity, Ref. [4], see Section 2.4 for furher discussions.

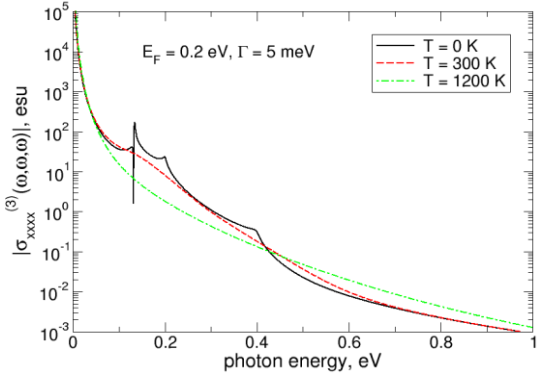

**Fig. 2.** The absolute value of the third conductivity $\sigma_{xxxx}^{(3)}(3\omega;\omega,\omega,\omega)$ of a single graphene layer as a function of the incident photon energy $\hbar\omega$ at the Fermi energy 0.2 eV and the effective relaxation rate 5 meV. Three resonant features at T=0 K correspond to the one-, two-, and three-photon absorption at the Fermi edge ($\hbar\omega=2E_F$, $E_F$ and $2E_F/3$).

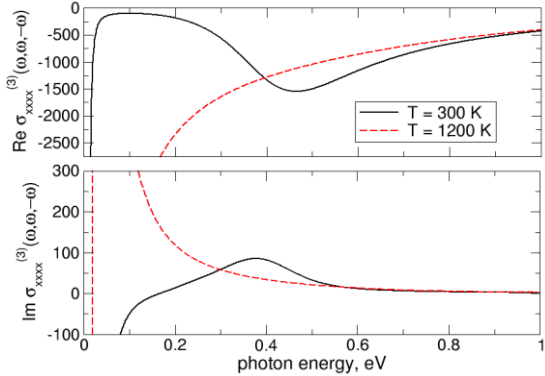

**Fig. 3.** The real and imaginary parts of the third conductivity $\sigma_{xxxx}^{(3)}(\omega;\omega,\omega,-\omega)$ as a function of $\hbar\omega$ at $E_F=$ 0.2 eV and $\Gamma$=5 meV.

*2.1.3. Second harmonic generation from graphene driven by a direct current*

Graphene is a centro-symmetric material, therefore under the action of a uniform external electromagnetic field it may produce only odd frequency harmonics. The central symmetry can be broken if a strong direct current is passed through the layer. Figure 4 shows the function $\sigma_{xxxx}^{(3)}(2\omega;\omega,\omega,0)$ which determines the second harmonic generation from a single graphene layer irradiated by an electromagnetic wave with the frequency $\omega$ and driven by a strong direct current. Two resonances at $\hbar\omega$=0.2 and 0.4 eV correspond to the inter-band transitions at $\hbar\omega=E_F$ and $2E_F$. It is also noticeable that the absolute value of the function $\sigma_{xxxx}^{(3)}(2\omega;\omega,\omega,0)$ responsible for the second harmonic generation is several orders of magnitude larger than the third harmonic generation function $\sigma_{xxxx}^{(3)}(3\omega;\omega,\omega,\omega)$.

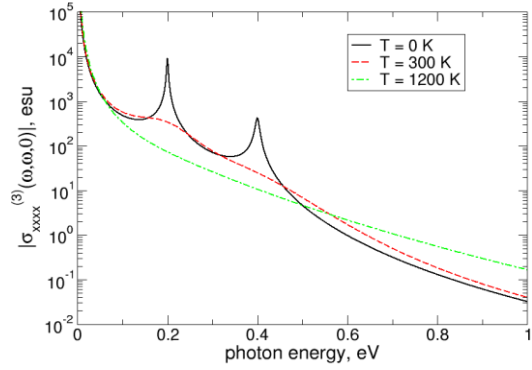

**Fig. 4.** The absolute value of the third conductivity $\sigma_{xxxx}^{(3)}(2\omega;\omega,\omega,0)$ of a single graphene layer as a function of $\hbar\omega$ at $E_F$=0.2 eV and $\Gamma$=5 meV. The resonant features at T=0 K correspond to $\hbar\omega=2E_F$ and $E_F$.

Similarly one can consider other nonlinear effects.

## 2.2. Influence of Substrates on the Third Harmonic Generation in Graphene

All results discussed so far refer to an isolated graphene layer in air. In real experiments graphene lies on a substrate. The influence of different types of substrates on the third harmonic generation efficiency was studied in Refs. [5,6]. Different types of substrates were considered and their very strong influence on the harmonic generation efficiency was demonstrated, Figure 5. The results can be summarized as follows.

*2.2.1. Dielectric dispersionless substrate*

If graphene lies on a dielectric layer of thickness $d$ with a frequency independent refractive index $n$ the third harmonic generation efficiency $\eta^{(3)}=I_{3\omega}/I_\omega^3$ can only be smaller or equal to the efficiency $\eta^{(3)}$ in isolated graphene. The substrate does not influence $\eta^{(3)}$ only if the substrate thickness $d$ is a multiple of $\lambda/2$, $2d/\lambda$ = integer, where $\lambda$ is the wavelength of radiation in the dielectric. If this condition is not satisfied the harmonic generation efficiency can be suppressed by *many orders of magnitude*, as compared to the isolated graphene layer.

*2.2.2. Polar dielectric substrate*

If the substrate is made out of a polar dielectric with one or a few optical phonons resonances in the dielectric function $\varepsilon(\omega)$, the efficiency $\eta^{(3)}$ can be substantially increased in certain frequency intervals,

Fig. 5(a), due to resonances at the TO and LO phonon frequencies. In addition, in the frequency range between $\omega_{TO}$ and $\omega_{LO}$ a flattening of the frequency dependence of $\eta^{(3)}$ can be observed, Fig. 5(a). This effect can be useful for some applications since in the isolated graphene the efficiency $\eta^{(3)}$ very quickly falls down with the frequency. The use of the polar dielectric substrate may help to achieve almost frequency independent third-order response of graphene. This effect is the case in the range of TO/LO phonon frequencies, i.e. at a few to a few tens of THz.

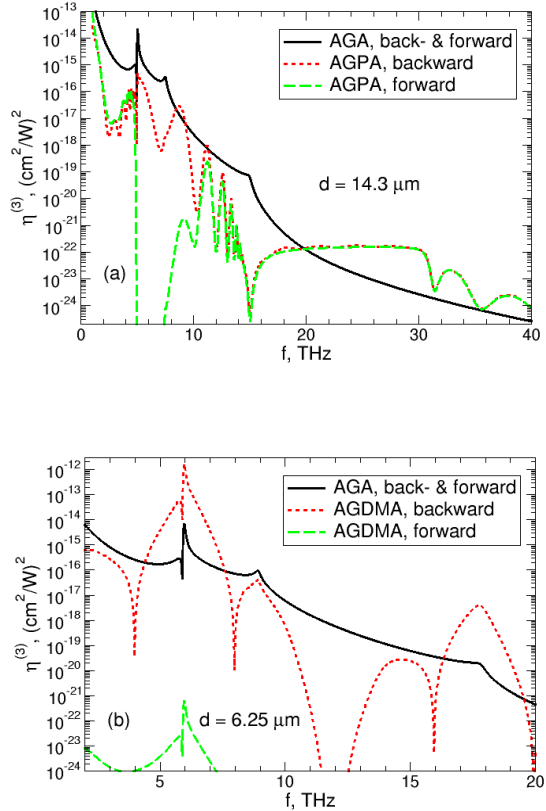

**Fig. 5.** Third harmonic generation efficiency $\eta^{(3)}=I_{3\omega}/I_\omega^3$ as a function of frequency: (a) in a structure air-graphene-polar dielectric-air (AGPA) and (b) air-graphene-dielectric-metal-air (AGDMA). Black curves show $\eta^{(3)}$ for isolated graphene (structure AGA) at T=0 K; resonances are seen at the frequencies $\hbar\omega=2E_F$, $E_F$ and $2E_F/3$. The transverse (TO) and longitudinal (LO) optical phonon frequencies in (a) lie at 15 and 30 THz respectively, $d$ is the dielectric thickness. The density of electrons is (a) $0.7\times10^{11}$ cm$^{-2}$ and (b) $10^{11}$ cm$^{-2}$.

*2.2.3. Dielectric dispersionless substrate with a metalized back side*

If the dielectric substrate with a frequency independent refractive index is covered by a thin metallic layer on the back side the efficiency of the third harmonic generation can be increased by *more than two orders of magnitude* as compared to the isolated graphene, Figure 5(b). This happens if $2d/\lambda$ = integer + 1/2 and is due to Fabry-Pérot resonances in the dielectric substrate.

The correct choice of the substrate material and thickness is thus very important for successful device operation. The substrates may both suppress and increase the third harmonic generation efficiency $\eta^{(3)}$ by several orders of magnitude.

**2.3. Nonperturbative Quasi-Classical Theory of the Nonlinear Effects in Graphene**

So far we discussed the results of the quantum theory of the third-order response of graphene which are valid at all frequencies ($\hbar\omega$ is smaller and larger than $2E_F$) but restricted by the third order Taylor expansion of the current in powers of the electric field, $j \sim \sigma^{(1)}E_0 + \sigma^{(3)}E_0^3$. At low (microwave, terahertz) frequencies, $\hbar\omega<2E_F$, one can develop a *nonperturbative* quasi-classical theory of the graphene response. At $\hbar\omega<2E_F$ the interband transitions from the valence to the conduction band can be neglected and the nonlinear graphene response can be described by quasi-classical Boltzmann equation. If the scattering integral is written in the $\tau$-approximation form ($\tau$ is the momentum relaxation time) and the external ac electric field is uniform, Boltzmann equation can be solved exactly [7].

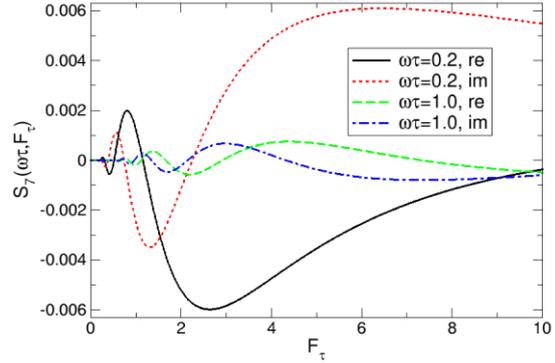

**Fig. 6.** A generalized conductivity $\sigma_{7\omega,\omega}(\omega\tau,F_\tau) \equiv \sigma_0 S_7(\omega\tau,F_\tau)$, which determines the 7th harmonic current $j_{7\omega}$ in response to the $\omega$-harmonic of the external ac field $E_\omega$, as a function of the field strength $F_\tau = eE_\omega\tau/p_F$; here $p_F$ is the Fermi momentum and $\sigma_0$ is the static Drude conductivity.

Using the results obtained in [7] one can calculate the generation efficiency of all (odd) frequency harmonics in arbitrarily strong electric fields. Figure 6 shows as an example the field dependence of the generalized complex conductivity of the graphene layer $\sigma_{n\omega,\omega}(\omega\tau,F_\tau)$, which is defined as the ratio of the $n$-th harmonic of the current to the 1st harmonic of the field (in Fig 6 $n=7$).

We have also calculated the transmission $T$, reflection $R$ and absorption $A$ coefficients of the wave with the frequency $\omega$ passing through a single graphene layer, at arbitrary values of the incident wave power. The frequency and power dependencies of the coefficients $T$, $R$ and $A$ also depend on the ratio $\gamma_{rad}/\gamma$ of the radiative decay rate $\gamma_{rad}$ to the dissipative damping rate $\gamma=1/\tau$ [7]. In all values of two parameters $\omega\tau$ and $\gamma_{rad}/\gamma$ the reflection coefficient decreases while the transmission coefficient grows with the wave power. The absorption coefficient $A$ typically decreases (the saturable absorption effect) but in some cases (e.g. at $\gamma_{rad}/\gamma \sim 2$) the absorption $A$ first grows and then decreases with power [7].

### 2.3. Influence of Nonlinearities on the Spectrum of Plasmons in Graphene

The spectrum of plasma waves $\omega=\omega_p(q)$ in two-dimensional (2D) electron systems is determined by the dispersion equation [8]

$$1+2\pi i\sigma(\omega)q/\omega\kappa=0, \qquad (2)$$

where $\omega$ and $q$ are the frequency and the wave vector of the 2D plasmon, $\sigma(\omega)$ is the dynamic conductivity of the 2D layer and $\kappa$ is the dielectric constant of the surrounding medium. The spectrum of 2D plasmons in graphene at $\hbar\omega<2E_F$ can be obtained from (2) if to use for $\sigma(\omega)$ the dynamic conductivity of graphene, see e.g. [9]. In the nonlinear regime, when the 2D plasmons are excited by a strong electromagnetic wave, the linear graphene conductivity $\sigma(\omega)$ should be replaced by the field dependent conductivity $\sigma_{\omega,\omega}(\omega\tau,F_\tau)$ derived in [7]. Using this way we have studied the influence of nonlinear effects on the frequency, linewidth, wavelength and propagation length of two-dimensional plasmons in graphene [10]. Figure 7 qualitatively illustrates the influence of the electric field strength on the far-infrared absorption spectrum in an array of narrow graphene stripes. A narrow weak-field plasma resonance centered at the plasma frequency $\omega_p$ shifts to lower frequencies and gets broader in the strong field regime. These predictions have been experimentally confirmed.

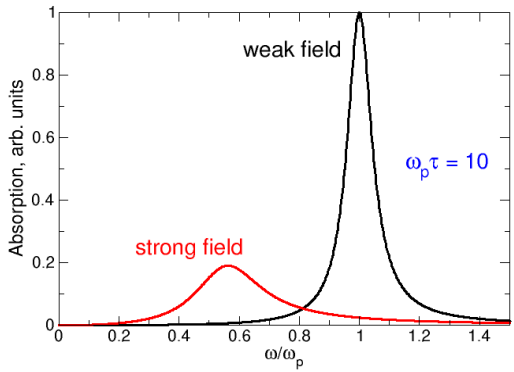

**Fig. 7.** Influence of the electric field strength on the shape of plasma resonances in a FIR absorption experiment in an array of narrow graphene stripes.

### 2.4. Experiments on the Nonlinear Graphene Electrodynamics

*2.4.1. Electrical control of the nonlinear response of graphene*

Experimentally nonlinear electrodynamic and optical effects in graphene have been studied since 2010, e.g. Refs. [11,12]. In early works the role of the parameter $\hbar\omega/2E_F$ which allows one to take advantage of the inter-band resonances was not fully understood and the experiments have been mainly performed on undoped graphene and at a few fixed incident wave frequencies. After the quantum theory of the nonlinear effects in graphene appeared (Ref. [3]) a number of experiments have been done aiming to use the unique graphene property to control its nonlinear response by the gate voltage. The electrical control of the four-wave mixing effect in graphene was first demonstrated in Ref. [13]. The gate voltage dependence of the third harmonic generation was demonstrated in the preprint [14]. These experiments confirmed the predictions of the theory [3].

*2.4.2. Analysis of the Optical Kerr Effect Measured by the Heterodyne Detection Technique*

One of the nonlinear phenomena potentially important for applications is the optical Kerr effect. It is typically measured by the Z-scan method or by using a more elaborated optical heterodyne detection (OHD) technique. In a recent paper [4] the optical Kerr effect has been studied by both techniques and the *effective* nonlinear refractive index $n_2$ of graphene was found to be about $-10^{-9}$ cm$^2$/W at the telecommunication wavelength $\sim 1.6$ μm.

In our recent paper [15] we have performed a detailed theoretical analysis of the OHD technique. First of all we have shown that the quantity $n_2$ which is typically used for characterization of the Kerr effect in *bulk* (three-dimensional) materials is *inapplicable* for two-dimensional crystals. Instead of the scalar real parameter $n_2$ one should use the complex fourth rank tensor $\sigma_{\alpha\beta\gamma\delta}^{(3)}(\omega;\omega,-\omega,\omega)$ to characterize the Kerr effect in atomically thin materials like graphene. We have shown how the quantites measured in a typical OHD experiment (e.g. in [4]) are related to the real and imaginary parts of the third conductivity tensor components and how to modify the experiment [4] in order to extract all nonlinear components of the tensor $\sigma_{\alpha\beta\gamma\delta}^{(3)}(\omega;\omega,-\omega,\omega)$ in dependence of the radiation frequency and polarization, charge carrier density, temperature and scattering parameters of graphene.

### 3. Summary

(1) Due to the linear energy dispersion of graphene electrons this material has strongly nonlinear electrodynamic and optical properties.

(2) The unique feature of graphene is that its nonlinear electrodynamic response strongly depends on the position of Fermi energy and hence on the electron density. This opens up opportunities to electrically control the nonlinear response of graphene by the gate voltage.

(3) The analytical quantum theory of the nonlinear electrodymanic response of graphene developed in recent years predicts a large number of interesting nonlinear effects many of which have not yet been experimentally studied. Available experimental results are in good agreement with the theory. Further experimental and theoretical studies of the nonlinear effects in graphene are needed.

(4) The unique nonlinear properties of graphene pave new ways to the development of novel electronic and photonic devices operating in the very broad frequency range from microwave and terahertz up to infrared and visible light frequencies.

## Acknowledgements

This work was funded by the European Union's Horizon 2020 research and innovation programmes Graphene Core 1 and Graphene Core 2 under Grant Agreements No. 696656 and No. 785219.

## References

[1]. A. K. Geim and K. S. Novoselov, The rise of graphene, *Nature materials*, Vol. 6, 2007, pp. 183-191.
[2]. S. A. Mikhailov, Non-linear electromagnetic response of graphene, *Europhysics Letters*, Vol. 79, 2007, 27002.
[3]. S. A. Mikhailov, Quantum theory of the third-order nonlinear electrodynamic effects of graphene, *Phys. Rev. B*, Vol. 93, 2016, 085403.
[4]. E. Dremetsika et al., Measuring the nonlinear refractive index of graphene using the optical Kerr effect method, *Optics Letters*, Vol. 41, 2016, 3281-3284.
[5]. N. A. Savostianova and S. A. Mikhailov, Giant enhancement of the third harmonic in graphene integrated in a layered structure, *Appl. Phys. Lett*, Vol. 107, 2015, 181104.
[6]. N. A. Savostianova and S. A. Mikhailov, Third harmonic generation from graphene lying on different substrates: optical-phonon resonances and interference effects, *Optics Express*, Vol. 25, 2017, pp. 3268-3285.
[7]. S. A. Mikhailov, Nonperturbative quasiclassical theory of the nonlinear electrodynamic response of graphene, *Phys. Rev. B*, Vol. 95, 2017, 085432.
[8]. F. Stern, Polarizability of a two-dimensional electron gas, *Phys. Rev. Lett.*, Vol. 18, 1967, pp. 546-548.
[9]. S. A. Mikhailov and K. Ziegler, New electromagnetic mode in graphene, *Phys. Rev. Lett.*, Vol. 99, 2007, 016803.
[10]. S. A. Mikhailov, Influence of optical nonlinearities on plasma waves in graphene, *ACS Photonics*, Vol. 4, 2017, pp. 3018-3022.
[11]. M. Dragoman et al., Millimeter-wave generation via frequency multiplication in graphene, *Appl. Phys. Lett.* Vol. 97, 2010, 093101.
[12]. E. Hendry, P. J. Hale, J. J. Moger, A. K. Savchenko, and S. A Mikhailov, Coherent Nonlinear Optical Response of Graphene, *Phys. Rev. Lett.,* Vol. 105, 2010 097401.
[13]. K. Alexander, N. A. Savostianova, S. A. Mikhailov, B. Kuyken, and D. Van Thourhout, Electrically tunable optical nonlinearities in graphene-covered SiN waveguides characterized by four-wave mixing, *ACS Photonics*, Vol. 4, 2017, 3039-3044.
[14]. G. Soavi et al., Broadband, electrically tuneable, third harmonic generation in graphene, arXiv:1710.03694.
[15]. N. A. Savostianova and S. A. Mikhailov, Optical Kerr Effect in Graphene: Theoretical Analysis of the Optical Heterodyne Detection Technique, *to be published*, arXiv:1801.09785.